\documentstyle[12pt]{article}
\advance\hoffset by -7mm
\makeatletter
\@addtoreset{equation}{section}
\makeatother

\renewcommand{\title}[1]{\null\vspace{25mm}

\noindent{\Large{\bf #1}}\vspace{10mm}

\noindent {\large By }}
\newcommand{\authors}[1]{\noindent{\large #1}\vspace{3mm}

}
\newcommand{\address}[1]{\noindent #1\vspace{5mm}

}
\renewcommand{\abstract}[1]{\vspace{19mm}

\noindent{\small{\em Abstract.} #1}\vspace{2mm}

}
\begin{document}
\rightline{ETH-TH/94-20,\quad hep-th/9411197}
\title{Normalization of scattering states and Levinson's\\[2mm]
theorem: reply to a comment by R. G. Newton\\[2mm]
[Helv. Phys. Acta {67}, 20 (1994)]}
\authors{Nathan Poliatzky}
\address{Institut f\"ur Theoretische Physik, ETH-H\"onggerberg,
CH-8093 Z\"urich, Switzerland\footnote{Electronic address: poli@itp.ethz.ch}}
\abstract{Normalization of scattering states, Levinson's theorem
and the high energy behavior of scattering phase shifts are discussed
in the light of R. G. Newton's recent criticism.}

\section{Introduction}

Recently R. G. Newton published a comment~\cite{Newton} criticizing
the methods and the results of a paper published in this
journal~\cite{polscatt}. His criticism touches on a few key points of
the subject and hence deserves a detailed reply. Here is the reply,
point by point, to his criticism.

\section{Normalization of scattering states}

Let us first consider the case without a threshold resonance
(i.\ e.\ either a finite solution with $k=0$ does not exist or it is a
bound state). We will return below to the (rather exceptional) case
with a threshold resonance (sometimes called a half-bound state or a
zero-energy solution). In this section we consider the case of a
Schr\"odinger equation; the case of a Dirac equation is completely
analogous and the relevant formulae can be found in~\cite{polscatt}.

The usual statement on the normalization of scattering states for a
central potential is
\begin{equation}\int_0^\infty{\rm d}r\;u_{kl}\left(r\right)u_{k'l}\left(r
\right)=2\pi\delta\left(k-k'\right)\label{norma}\end{equation}
The question then arises of what
happens in the limit $k\rightarrow k'$. Of course the right-hand side
becomes infinite, but it is easy to see that the difference, between
this equation and its free-particle counterpart, is finite.
In~\cite{polscatt} it was proven that
\begin{equation}\int_0^\infty{\rm d}r\;\left[u_{kl}^2\left(r\right)-v_{kl}^2
\left(r\right)\right]=2\eta'_l\left(k\right)\label{diagn}\end{equation}
where $v_{kl}\left(r\right)$  are the free-particle wave functions and
$\eta_l\left(k\right)$ is the phase shift.

Equations~(\ref{norma}) and~(\ref{diagn}) can be put together as
one compact equation
\begin{equation}\int_0^\infty{\rm d}r\;u_{kl}\left(r\right)u_{k'l}\left(r
\right)=2\pi\delta\left(k-k'\right)+2\eta'_l\left(k\right)\delta_{kk'}
\label{normb}\end{equation}
where $\delta_{kk'}=1$ if $k=k'$ and $\delta_{kk'}=0$ otherwise.

Equation~(\ref{normb}) is what R. G. Newton criticizes. As he put it:
``He arrives at a modification of the usual quasi-orthogonality of the
scattering wave functions in which the Dirac distribution is
supplemented by a bounded function that differs from zero at only one
point. In any context in which the Dirac distribution is meaningful,
such a function is of course equal to zero, and its addition is of no
consequence.''

Admittedly, if the delta function on the right-hand side
of~(\ref{normb}) is interpreted strictly as a distribution, the
addition of the second term is of no consequence and
equations~(\ref{normb}) and~(\ref{norma}) are identical. However, the
above criticism misses several points. One is that~(\ref{normb}) is
merely a short-hand description of the two
equations~(\ref{norma}) and~(\ref{diagn}), the validity of which is
beyond any doubt. So that, if needed, any reference to~(\ref{normb})
can be avoided.

A deeper point is that the Dirac delta function and a distribution are
not exactly identical concepts, e.\ g.\ the delta function on the
right-hand side of~(\ref{norma}) or~(\ref{normb}) cannot be
interpreted as a distribution in the usual mathematical sense. The
proof of this statement is as follows. Assume that this delta function
is a distribution. However, for the left-hand side of~(\ref{norma})
it perfectly makes sense to subtract its free-particle counterpart and
then put $k=k'$. The result is finite and given in~(\ref{diagn}). On
the other hand, for the right-hand side of~(\ref{norma}) such an
operation does not make sense (or its outcome is ambiguous), if the
delta function is interpreted as a distribution. Hence the left- and
the right-hand side of~(\ref{norma}) cannot be the same object and
the conclusion is: the right-hand side of equation~(\ref{normb})
is a modification of the usual notion of a distribution.

The mathematical
notion of a distribution is based on a class of all possible limiting
processes, whereas in quantum theory the Dirac delta function is
always either a particular limiting process or a particular class of
limiting processes with a fixed physical interpretation of the
parameters involved.\footnote{For instance, in the usual calculation
of a differential cross section in quantum field theory one has to
square a Dirac delta function, which is done unambiguously using a
particular physically motivated limiting process. From the point of
view of a distribution this does not make sense.}  Having this in mind
there is no danger of ambiguities for the use of equation~(\ref{normb}).

\section{Levinson's theorem}\label{levohne}

Consider the completeness relation
\begin{equation}\sum_{\epsilon_l\leq0}u_{\epsilon_ll}\left(r\right)
u_{\epsilon_ll}
\left(r'\right)+\int_0^\infty\frac{{\rm d}k}{2\pi}\,u_{kl}\left(r\right)u_{kl}
\left(r'\right)=\delta\left(r-r'\right)\label{compl}\end{equation}
where the sum runs over the bound and the integral over the
scattering states.

If the right-hand side of~(\ref{compl}) is interpreted as a Dirac
delta function (i.\ e.\ as a class of limiting processes which are
independent of the potential) and not as a usual distribution, we can
subtract from~(\ref{compl}) its free particle counterpart and put
$r=r'$. The delta functions cancel and we obtain
\begin{equation}\sum_{\epsilon_l\leq0}u_{\epsilon_ll}^2\left(r\right)+
\int_{0}^\infty\frac{{\rm d}k}{2\pi}\,\left[u_{kl}^2\left(r\right)-v_{kl}^2
\left(r\right)\right]=0\label{complb}\end{equation}
Integrating~(\ref{complb}) over $r$ and substituting~(\ref{diagn}), we
obtain the Levinson theorem
\begin{equation}\eta_l\left(0\right)=n_l\pi\label{Levinf}\end{equation}
for the case of a Schr\"odinger equation, where $n_l$ is the number of
bound states. If the same is carried out for the case of a Dirac
equation (see ref.~\cite{polscatt} for the details), the result is
\begin{equation}\eta_{m\kappa}\left(0\right)+\eta_{-m,\kappa}\left(0\right)=
\left(N_\kappa^++N_\kappa^-\right)\pi\label{Levd}\end{equation}
where $N_\kappa^+$ is the number of positive and $N_\kappa^-$ is the
number of negative energy bound states.\footnote{More precisely we obtain
\[\eta_l\left(0\right)=\eta_l\left(\infty\right)+n_l\pi\]
for the case of a Schr\"odinger equation, and
\[\eta_{m\kappa}\left(0\right)+\eta_{-m,\kappa}\left(0\right)=
\eta_{\infty\kappa}\left(\infty\right)+\eta_{-\infty,\kappa}
\left(\infty\right)+\left(N_\kappa^++N_\kappa^-\right)\pi\]
for the case of a Dirac equation. However, we are free to put
$\eta_l\left(\infty\right)=0$ and $\eta_{\infty\kappa}\left(\infty\right)
+\eta_{-\infty,\kappa}\left(\infty\right)=0$.}

Looking at~(\ref{Levd}) it is natural to ask whether a stronger
statement of Levinson's theorem is possible, valid for the positive
and negative energy phase shifts separately. It turns out to be the
case~\cite{pollev}. The clue to derive such a stronger statement of
Levinson's theorem is simple: consider the second order (iterated)
Dirac equation and take the limit $k\rightarrow0$. In this
nonrelativistic limit\footnote{It is an interesting fact that such a
non-relativistic limit which, from the physical point of view, is a
most natural one, does not coincide with the usual expansion based on
the Foldy-Wouthuysen scheme (see~\cite{FW}).} the second order Dirac
equation becomes identical to a Schr\"odinger equation, so that at
$k=0$ one can relate the phase shifts of a Dirac equation to the phase
shifts of a corresponding Schr\"odinger equation. The stronger
statement of Levinson's theorem obtained in this way is:
\begin{equation}\eta_{m\kappa}\left(0\right)=n_\kappa^+\pi
\label{Levwdsolpintr}\end{equation}
\begin{equation}\eta_{-m\kappa}\left(0\right)=n_\kappa^-\pi
\label{Levwdsolmintr}\end{equation}
where $n_\kappa^+$ is the number of positive and $n_\kappa^-$ the
number of negative energy nodes of the $k=0$ solution of the Dirac
equation. Unlike the case of a Schr\"odinger equation, there is a
subtlety here: for a given energy sign the number of bound states may
not equal the number of nodes of a $k=0$ solution, i.\ e.\ in general
$N_\kappa^+\neq n_\kappa^+$ and $N_\kappa^-\neq n_\kappa^-$. The
equality holds only for the total numbers:
\begin{equation}N_\kappa^++N_\kappa^-=n_\kappa^++n_\kappa^-
\label{Nnpm}\end{equation}

The generality and simplicity of the above proof of Levinson's theorem
speaks for itself, and can be viewed as a confirmation of the validity
and usefulness of the above discussion on the normalization of
scattering states. Notice that we did not have to assume anything
explicit about the potential. The only implicit assumption is that the
potential decays faster than $1/r^2$ at large distances, which is
sufficient for the validity of~(\ref{diagn}).

However, when in ref.~\cite{polscatt} this proof was extended to the
case with a threshold resonance, a surprising result was obtained:
\begin{equation}\eta_l\left(0\right)=\left(n_l+q\right)\pi
\label{Levinfa}\end{equation}
in the case of a Schr\"odinger equation, and
\begin{equation}\eta_{\pm m\kappa}\left(0\right)=\left(n_\kappa^\pm
+q\right)\pi\label{Levwdsolp}\end{equation}
in the case of a Dirac equation, where either $q=0$, or $q=\frac{1}{4}$,
or $q=\frac{1}{2}$. For the case with a threshold resonance $q=0$ can
be ruled out and $q=\frac{1}{2}$ is what one would have expected.
The surprise is in the possibility of allowing for $q=\frac{1}{4}$,
which is the subject of R. G. Newton's criticism. From the point of
view of a conventional proof of Levinson's theorem, where there is
always a certain restriction on the potential, the case $q=\frac{1}{4}$
cannot be ruled out completely. However while the
paper~\cite{polscatt} was in press it was found that the derivation of
the normalization integral actually implies a certain constraint which
was overlooked in~\cite{polscatt} and which rules out the case
$q=\frac{1}{4}$. This is the reason why in a subsequent
paper~\cite{pollev} the case $q=\frac{1}{4}$ was not mentioned.
Meanwhile M. Sassoli de Bianchi~\cite{sassoli} investigated the
problem within the framework of one spatial dimension and came to
the same conclusion. In the following section these matters are
considered in more detail.

\section{The case with a  threshold resonance}

Assume that the coupling constant is tuned such that a finite,
$k=0$ solution (threshold resonance) exists. By examining its
asymptotic behavior at large distances, it is easily seen that such a
solution is normalizable (i.e. a bound state), unless $l=0$, in the
case of a Schr\"odinger equation, and $\kappa=-1,\;\epsilon=m$ or
$\kappa=1,\;\epsilon=-m$, in the case of a Dirac
equation.\footnote{Notice that, for a given angular momentum, at most
one threshold resonance is possible, either \mbox{with $\epsilon=m$ or
with $\epsilon=-m$}.} Hence in this section $l=0$, in the case of a
Schr\"odinger equation, and $\kappa=\pm1$, in the case of a Dirac
equation.

According to~\cite{polscatt} in the presence of a threshold
resonance~(\ref{diagn}) becomes modified according to
\begin{equation}\int_0^\infty{\rm d}r\;\left[u_{k0}^2\left(r\right)-v_{k0}^2
\left(r\right)\right]=2\eta'_0\left(k\right)+2\pi\delta\left(k\right)
\sin^2\eta_0\left(k\right)\label{diagna}\end{equation}

As in section~\ref{levohne}, integrating~(\ref{complb}) over $r$ and
substituting~(\ref{diagna}), we obtain Levinson's theorem for the case
with a threshold resonance
\begin{equation}\eta_0\left(0\right)=n_0\pi+\frac{\pi}{2}\sin^2\eta_0\left(0
\right)\label{Levmod}\end{equation}
Equations~(\ref{diagna}) and~(\ref{Levmod}) are valid for a
Schr\"odinger equation. In the case of a Dirac equation, as was mentioned
in the preceding section, one can rewrite the Dirac equation at $k=0$
as an ordinary Schr\"odinger equation and use~(\ref{Levmod}) to obtain
the stronger statement of Levinson's theorem for a Dirac
equation:
\begin{equation}\eta_{\pm m\kappa}\left(0\right)=n_\kappa^\pm\pi
+\frac{\pi}{2}\sin^2\eta_{\pm m\kappa}\left(0\right)
\label{Levmoddir}\end{equation}
where $n_\kappa^+$ is the number of positive and $n_\kappa^-$ the
number of negative energy nodes of the $k=0$ solution.

Equation~(\ref{Levmod}) was first derived in ref.~\cite{Ni}. The authors
did not check for all solutions of this equation but instead were
content by verifying that~(\ref{Levinfa}) for $q=0$ and
$q=\frac{1}{2}$ is a solution of~(\ref{Levmod}).
Equations~(\ref{Levmod}) and~(\ref{Levmoddir}) were derived
independently in ref.~\cite{polscatt} and it was pointed out that
there is a third solution with $q=\frac{1}{4}$. However it was not
noticed that the derivation of~(\ref{diagna}) also implies
\begin{equation}\sin\left[2\eta_l\left(0\right)\right]=0
\label{constr}\end{equation}
This constraint rules out the case $q=\frac{1}{4}$
in~(\ref{Levinfa}).\footnote{The situation for a Dirac equation is
completely analogous and the constraint is
\[\sin\left[2\eta_{\pm m\kappa}\left(0\right)\right]=0\]
This constraint rules out the case $q=\frac{1}{4}$ in~(\ref{Levwdsolp}).}

To understand how this constraint comes about, we now consider the
derivation of~(\ref{diagna}) in more detail (section 2.1 of
ref.~\cite{polscatt}). We start with the exact identity
\begin{equation}\int_0^R{\rm d}r\;u_{kl}^2\left(r\right)=\frac{1}{2k}\left[
u'_{kl}\left(R\right)\partial_ku_{kl}\left(R\right)-u_{kl}\left(R\right)
\partial_ku'_{kl}\left(R\right)\right]\label{Rnormb}\end{equation}
and assume that the potential $V\left(r\right)$ vanishes for $r\geq R$.
Then the right-hand side of~(\ref{Rnormb}) depends only on the
asymptotic wave functions and thus can be calculated. The result is
\begin{equation}\int_0^R{\rm d}r\;u_{kl}^2\left(r\right)=2R+2\eta'_l\left(k
\right)-\left(-1\right)^l\frac{1}{k}\,\sin\left[2kR+2\eta_l\left(k\right)
\right]\label{Rnormc}\end{equation}
Subtracting from~(\ref{Rnormc}) its free-particle counterpart and
expanding the sinus, we obtain

\begin{equation}\int_0^R{\rm d}r\;\left[u_{kl}^2\left(r\right)-
v_{kl}^2\left(r\right)\right]=2\eta'_l\left(k\right)+\left(-1\right)^l
\left\{2\sin^2\eta_l\left(k\right)\frac{\sin\left(2kR\right)}{k}
-\frac{\sin\left[2\eta_l\left(k\right)\right]}{k}\cos\left(2kR\right)\right\}
\label{Rnormca}\end{equation}
The second term in curly parenthesis possesses a $k^{-1}$ singularity
unless $\sin\left[2\eta_l\left(0\right)\right]=0$. In the case with a
threshold resonance, where the $k=0$ state is part
of scattering states, such a singularity is intolerable (for instance,
the integral over $k$ in~(\ref{complb}) includes the point $k=0$ in
this case). Hence the constraint~(\ref{constr}) must be satisfied. The
derivation of~(\ref{diagna}) is complete now, if one sends
$R\rightarrow\infty$.

\section{High energy limit}

In~\cite{polscatt} it was shown that the high energy behavior of the
scattering phase shifts is
\def\ar#1{\hbox to 0pt{\lower 5.0pt\hbox{\hskip4pt$\scriptstyle#1\rightarrow
            \infty$}\hss}\hbox to 30pt{\rightarrowfill}}
\begin{equation}\eta_l\left(k\right)\;\ar k\;k\int_0^\infty{\rm d}r\;\left[
\sqrt{1-\frac{2m}{ k^2}V\left(r\right)}-1\right]\,,
\label{etaklargea}\end{equation}
in the case of a Schr\"odinger equation, and
\begin{equation}\eta_{\epsilon\kappa}\left(k\right)\;\ar k\;k\int_0^\infty{\rm
d}r\;\left[\sqrt{1-\frac{2\epsilon}{ k^2}V\left(r\right)}-1\right]\,,
\label{etakdlargea}\end{equation}
in the case of a Dirac equation. Obviously, if the potential $V$ is
less singular than $1/r$ at the origin, then~(\ref{etaklargea})
and~(\ref{etakdlargea}) lead to the well known results
\begin{equation}\eta_l\left(\infty\right)=0\label{etainf}\end{equation}
in the case of a Schr\"odinger equation, and
\begin{equation}\eta_{\epsilon\kappa}\left(k\right)\;\ar k\;
=-\frac{\epsilon}{\vert\epsilon\vert}\int_0^\infty{\rm d}r\;
V\left(r\right)\label{etakdlarge}\end{equation}
in the case of a Dirac equation.

R. G. Newton doubts the validity of~(\ref{etaklargea})
and~(\ref{etakdlargea}): ``His change of variables leads, without
comment, to a highly singular equation on a complex contour, and it is
not clear whether his manipulations are valid.''

This criticism is without foundation, since the equations,
which were used to derive~(\ref{etaklargea}) and~(\ref{etakdlargea}),
become singular only at certain finite values of $k$. Taking into
account the fact that these equations are local in $k$, such singular
behavior is irrelevant to the high energy limit $k\rightarrow\infty$.

\end{document}